\documentclass[%
 reprint,
superscriptaddress,
 amsmath,amssymb,
aps,
prb,
]{revtex4-1}
\usepackage{xfrac}
\usepackage{nicefrac}
\usepackage{graphicx}
\usepackage{dcolumn}
\usepackage{bm}
\usepackage{cleveref}
\usepackage{color}
\newcommand{\sample}{Ba(Fe$_{0.975}$Co$_{0.025}$)$_2$As$_{2}$ }
\newcommand{\samplens}{Ba(Fe$_{0.975}$Co$_{0.025}$)$_2$As$_{2}$}
\newcommand{\refns}{Ba(Fe$_{0.937}$Co$_{0.063}$)$_2$As$_{2}$}

\newcommand{\aogs}{$A_{1g}$ }
\newcommand{\bogs}{$B_{1g}$ }

\newcommand{\ohead}[1]{}

\newcommand{\neel}{Ne\'{e}l }

\begin{document}

\title{Elastoresistive and Elastocaloric Anomalies at Magnetic and Electronic-Nematic Critical Points}

\author{Alexander. T. Hristov}
\thanks{These authors contributed equally to this work.}
\affiliation{Geballe Laboratory for Advanced Materials, Stanford University, Stanford, CA 94305}
\affiliation{Stanford Institute for Materials and Energy Science, SLAC National Accelerator Laboratory, 2575  Sand  Hill  Road,  Menlo  Park,  California  94025,  USA}
\affiliation{Department of Physics, Stanford University, Stanford, CA 94305}

\author{Matthias S. Ikeda}
\thanks{These authors contributed equally to this work.}
\affiliation{Geballe Laboratory for Advanced Materials, Stanford University, Stanford, CA 94305}
\affiliation{Stanford Institute for Materials and Energy Science, SLAC National Accelerator Laboratory, 2575  Sand  Hill  Road,  Menlo  Park,  California  94025,  USA}
\affiliation{Department of Applied Physics, Stanford University, Stanford, CA 94305}

\author{Johanna C. Palmstrom}
\affiliation{Geballe Laboratory for Advanced Materials, Stanford University, Stanford, CA 94305}
\affiliation{Stanford Institute for Materials and Energy Science, SLAC National Accelerator Laboratory, 2575  Sand  Hill  Road,  Menlo  Park,  California  94025,  USA}
\affiliation{Department of Applied Physics, Stanford University, Stanford, CA 94305}

\author{Philip Walmsley}
\affiliation{Geballe Laboratory for Advanced Materials, Stanford University, Stanford, CA 94305}
\affiliation{Stanford Institute for Materials and Energy Science, SLAC National Accelerator Laboratory, 2575  Sand  Hill  Road,  Menlo  Park,  California  94025,  USA}
\affiliation{Department of Applied Physics, Stanford University, Stanford, CA 94305}

\author{Ian R. Fisher}
\affiliation{Geballe Laboratory for Advanced Materials, Stanford University, Stanford, CA 94305}
\affiliation{Stanford Institute for Materials and Energy Science, SLAC National Accelerator Laboratory, 2575  Sand  Hill  Road,  Menlo  Park,  California  94025,  USA}
\affiliation{Department of Applied Physics, Stanford University, Stanford, CA 94305}

\date{\today}

\begin{abstract}
Using \sample as an exemplar material exhibiting second order electronic-nematic and antiferromagnetic transitions, we present measurements that reveal 
anomalies in the elastoresistance ($\nicefrac{\partial \rho_{ij}}{\partial\varepsilon_{kl}}$) and elastocaloric effect ($\nicefrac{\partial T}{\partial\varepsilon_{kl}}$) at both phase transitions.
Both effects are understood to arise from the effect of strain on the transition temperatures; in the region close to the phase transitions this leads to (1) similarity between the strain and temperature derivatives of the resistivity and (2) similarity between the elastocaloric effect and the singular part of the specific heat.
These mechanisms for elastoresistance and elastocaloric effect should be anticipated for any material in which mechanical deformation changes the transition temperature.
Furthermore, these measurements provide evidence that the Fisher-Langer relation $\rho^{(c)} \propto U^{(c)}$ between the scattering from critical degrees of freedom and  their energy-density, respectively, holds near each of the Ne\'{e}l  and electronic nematic transitions in the material studied.

\end{abstract}

\maketitle

\ohead{Introduction to the Result of Fisher Langer}


In metals, critical anomalies in the resistivity and specific heat are widely studied because the scattering rate and density of states at the Fermi surface are sensitive to symmetry breaking order and its critical fluctuations, regardless of the symmetry broken by the order parameter.
Perhaps the best known theory connecting resistivity 
and specific heat in a critical system originates from a study by Fisher and Langer. \cite{FisherResistive1968}
The result of their work was to relate the contribution to the resistivity $\rho^{(c)}$ from critical fluctuations in a ferromagnet to the energy-density $U^{(c)}$ of the critical magnetic degrees of freedom. 
When this relation is satisfied, the change in the temperature derivative of the resistivity, $\frac{d\rho}{dT}$, is proportional to the critical contribution to the specific heat of a material, $C_p$.
Though it was originally postulated that the correspondence held only above the magnetic transition temperature, $T_c$, subsequent work has clarified conditions under which the relation holds both above and below $T_c$ and in both ferromagnets and antiferromagnets,\cite{AlexanderCritical1976} as well as applied the Fisher-Langer relation to nonmagnetic systems. \cite{SimonSpecific1971}

\begin{figure}
\includegraphics[width=\linewidth]{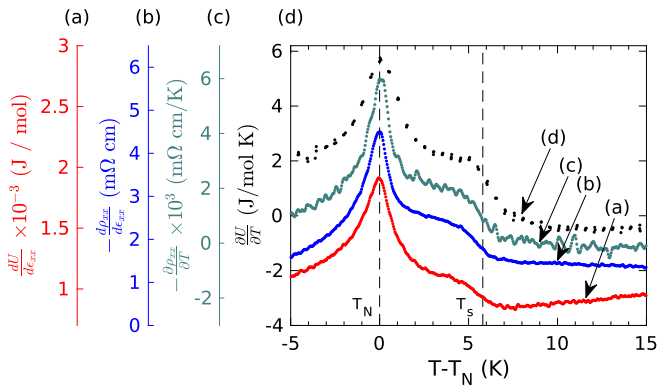}
\caption{(a-d) Temperature and strain derivatives of the resistance and energy in \sample in proximity to the successive nematic and \neel transitions, each plotted against a single corresponding axis on the left. Traces (a), (b), and (c) are obtained from a single sample with $T_N\approx 93$K. The specific heat shown in (d) is obtained from a second sample used in Ref.~\onlinecite{Chu_Determination_2009}. \label{fig:all}}
\end{figure}

Here, we study the effect of strain, $\varepsilon_{ij}$, on the same quantities studied by Fisher and Langer, the resistivity and energy.
\footnote{The strain derivative of the resistivity, $\nicefrac{\partial \rho_{ij}}{\partial\varepsilon_{kl}}$, is measured directly, and the strain derivative of $U^{(c)}$ is obtained from the product of the total specific heat (of which the critical contribution is less than 5~\%) and the elastocalioric effect, $\nicefrac{\partial T}{\partial\varepsilon_{kl}}$.}
Strain-based experimental techniques are currently utilized in a variety of contexts, including scattering,\cite{Chen2016,Dhital2014,Tam2017} transport\cite{chu2012divergent,kuo2013measurement,HicksPiezoelectric2014,IkedaSymmetric2018} and NMR measurements.\cite{KissikovNMRProbe2017,KissikovSusceptibility2017,kissikov2018uniaxial} 
In materials such as \samplens, where there is a second order electronic nematic transition,\citep{ni2008effects,Chu_Determination_2009,lester2009neutron, chu2012divergent} antisymmetric strain of the same symmetry as the order parameter has been used to measure the nematic susceptibility.\cite{chu2012divergent,kuo2013measurement,ShapiroSymmetry2015}
Here, rather than using strain of the same symmetry as the order parameters, we induce strain belonging to different irreducible representations from the order parameter. 
The distinction between the symmetric and broken symmetry phases remains sharp even in the presence of such strain, although the critical temperatures, $T_S$ and $T_N$, generically shift.
Close to criticality, this results in proportionality between the elastocaloric coefficient and $C_p$, and between elastoresistance and $\frac{d\rho}{dT}$.
This correspondence arises naturally from the strain dependence of the transition temperatures.
Furthermore, our results suggest the applicability of the Fisher-Langer relation in \samplens, giving rise to the generalized correspondence,
\begin{equation}
C_p \propto \frac{\partial\rho}{\partial T} \propto \frac{dU}{d\varepsilon_{kl}} \propto \frac{d\rho}{d\varepsilon_{kl}}, \label{eq:all}
\end{equation}
which holds near each phase transition.
\Cref{eq:all} is the key result of this work.  
It is manifest in \cref{fig:all}, in which the temperature dependences of these four separately measured quantities are shown to be strikingly similar.

\sample is an under-doped iron-pnictide with two separate continuous phase transitions upon cooling.\citep{ni2008effects,Chu_Determination_2009,lester2009neutron}
The first of these is an electronically-driven structural transition at $T_S\approx 99$~K in which the system changes from a tetragonal crystal structure to an orthorhombic structure through the onset of a nematic order parameter, which has $B_{2g}$ symmetry and couples to $\varepsilon_{xy}=\varepsilon_{B_{2g}}$ strain\cite{Fernandes_Scaling_2013,chu2012divergent}.
Additionally there is an antiferromagnetic transition at $T_N\approx 93$~K.
Due to the presence of spin orbit coupling, the antiferromagnetic polarization is along the nematic easy axis,\cite{kissikov2018uniaxial} so both transitions are of an Ising class. 
The antiferromagnetic phase transition is mediated by short range interactions for which the upper critical dimension $d_c^+ = 4$, whereas the nematic transition is mediated by long range strain forces, with $d_c^+=2$.\cite{Karahasanovic2016}

The samples of \sample were obtained by the same self-flux method described previously.\cite{Chu_Determination_2009}
For studies of the elastocaloric effect, reference specific heat data on a crystal of \sample (with $T_N\approx 92$~K and $T_S\approx 98$~K) were obtained from ref. \onlinecite{Chu_Determination_2009}, which used a standard relaxation technique in a Quantum Design Physical Property Measurement System (PPMS).
Measurements of resistivity, elastoresistivity, and elastocaloric effect were obtained simultaneously from a second sample ($T_N\approx 93$~K and $T_S\approx 99$~K), which has been measured previously to obtain precise values of the strain induced changes in the nematic and Neel transition temperatures.\cite{IkedaSymmetric2018}
The elastoresistance sample was cut into a bar with a long axis oriented with the $[100]$ axis of the tetragonal unit cell, with approximate dimensions $2$~mm  $\times$  $0.4$~mm $\times$ $0.035$~mm.\cite{IkedaSymmetric2018}
Uniaxial stress was applied to the sample using a commercially available strain apparatus (CS100, Razorbill Instruments), which almost perfectly compensates the thermal expansion of \samplens. \cite{IkedaSymmetric2018}
This experimental configuration allows the sample to be tuned near zero total strain at all temperatures.
The displacement-per-volt of the piezoelectric stacks was characterized by previous measurements \cite{IkedaSymmetric2018} using an Andeen Hagerling AH2550A capacitance bridge for this device; the displacement-per-volt of the PZT is frequency independent to a few percent.\cite{HristovMeasurement2018}
The piezo devices were driven in the range between 1Hz and 30Hz, with a maximum oscillatory amplitude of 7.5~V.
Four-point electrical contact was made using Chipquik SMD291AX10T5 solder and gold wires.
The voltage contacts were $0.65$~mm apart, and the crystal was mounted with epoxy so that a middle portion the crystal, measuring $0.780$~mm long and containing both voltage contacts, was between the mounting plates and experienced a piezo-driven mechanical deformation.
For the resistivity measurements, the temperature derivative was obtained from a smoothing-spline fit to the raw resistivity data.

The application of a uniaxial stress $\sigma_{xx}$ along the [100] axis induces a finite strain along all three crystal axes, as is evident by solving $\sum_{kl}c_{ijkl}\varepsilon_{kl}=\sigma_{ij}$, where $c_{ijkl}$ is the elastic constant tensor and $\sigma_{ij}$ is the stress. 
The combination of strains can be written, using compact Voigt notation, in a basis corresponding to irreducible representations of $D_{4h}$:
\begin{equation}
\varepsilon_{A_{1g},1} = \frac{1}{2} (\varepsilon_{xx}+\varepsilon_{yy}) = {\frac{\sigma_{xx}}{2(c_{11}+c_{12}-2\frac{c_{13}^2}{c_{33}})}} \label{astrain}
\end{equation}
\begin{equation}
\varepsilon_{A_{1g},2}={\varepsilon_{zz}}={\frac{\sigma_{xx}}{2c_{13}-(c_{11}+c_{12})\frac{c_{33}}{c_{13}}}}\label{zstrain}
\end{equation}
\begin{equation}
\varepsilon_{B_{1g}} = {\frac{1}{2} (\varepsilon_{xx}-\varepsilon_{yy})}  ={\frac{\sigma_{xx}}{2(c_{11}-c_{12})}\label{bstrain}}
\end{equation}
For the specific case of this sample and uniaxial stress along the [100] axis, the elastic constants in \cref{astrain,zstrain,bstrain} are only weakly temperature dependent and lack singular temperature dependence at criticality. 
It is appropriate to approximate them as temperature-independent constants for a narrow range of temperatures in the vicinity of the critical temperatures.
We emphasize, however, that this may not be appropriate (i) over a wider range of temperatures or (ii) for \sample samples under $B_{2g}$ strain, where the shear modulus $c_{66}$ softens due to coupling to order parameter fluctuations.\citep{Fernandes_Scaling_2013}

The strains induced by uniaxial stress, shown in \cref{astrain,zstrain,bstrain}, all preserve the primary mirror planes of the $D_{4h}$ point group; they belong to distinct irreducible representations from the nematic and \neel order parameters and so do not couple bilinearly to either order parameter.\cite{*[{The case of a transport measurement under an induced strain with the the same symmetry as the order parameter has been treated in }] [{ and in the experiments referenced therein.}] ShapiroSymmetry2015}
Generally, a strain $\varepsilon_\alpha$, belonging to an irreducible representation $\alpha$, couples to (one of several) order parameters $\psi_i$ belonging to different irreducible representation through terms in the Landau free energy that depend on second order moments of $\psi_i$.
Under such an induced strain, a primary effect of strain is to cause the transition temperature $T_i$ to vary according to
\begin{equation}
T_i =  T_{i,0} - \Lambda_{\alpha,i}^{(1)}\varepsilon_{\alpha} - \Lambda_{\alpha,i}^{(2)}\left(\varepsilon_{\alpha}\right)^2 +\ldots,\label{tc}
\end{equation}
where, for \sample, $i=S$ for the nematic/structural phase transition and, and $i=N$ for the \neel transition.
In the case $\varepsilon_\alpha$ breaks any crystal symmetries, $\Lambda^{(1)}_{\alpha,i}=0$; $\Lambda^{(1)}_{\alpha,i}\neq 0$ only for `trivial' strains which break no additional crystal symmetries, such as $\varepsilon_{zz}$ for a tetragonal material. We therefore phrase the discussion in terms of this general form (\cref{tc}) which simultaneously includes the case both when $\alpha$ is trivial and non-trivial. 
\footnote{In the limit of infinitesimal strains, one therefore anticipates that the linear variation from the \aogs strains will dominate over the quadratic contributions.}
We noted earlier that for a sample aligned along the principal tetragonal axes, multiple strains are applied simultaneously and are all in approximately fixed proportion to each other and $\varepsilon_{xx}$. 
To treat multiple simultaneous strains of fixed proportion, we define $\nu_{xx}^\alpha = \varepsilon_\alpha/\varepsilon_{xx}$ for each irreducible representation and replace \cref{tc} by a sum over the $\alpha$, which yields 
$\frac{dT_i}{d\varepsilon_{xx}}= \sum_\alpha \nu_{xx}^\alpha\Lambda_{\alpha,i}^{(1)}$ and $\frac{d^2T_i}{d\varepsilon_{xx}^2}= \sum_\alpha (\nu_{xx}^\alpha)^2\Lambda_{\alpha,i}^{(2)}$.
For the material studied here, these coefficients have been reported in ref. \onlinecite{IkedaSymmetric2018} for each of the two phase transitions; the linear response being given by $\frac{dT_N}{d\varepsilon_{xx}} = -629 \pm 2$ K, and $
\frac{dT_S}{d\varepsilon_{xx}} = -521 \pm 4$~\text{K}.
We note that there is no general (symmetry) reason for these two quantities to be within a factor of 2 -- it presumably reflects the common electronic physics driving the magnetic and the structural transitions.  
However, it greatly simplifies the further analysis, in that the strain-induced changes in $T_N$ and $T_S$ give rise to elastocaloric and elastoresistive anomalies of comparable magnitude at the two transitions.


\ohead{summary of elastoresistance technique}

The elastoresistance of the sample was measured using an amplitude demodulation technique described previously.\cite{HristovMeasurement2018} 
An AC strain is induced in the sample, which produces a time-varying resistivity that modulates the voltage from an AC current through the sample. 
This produces sideband signals with amplitude proportional to the elastoresistivity. 
Here, the sideband signal is directly detected using a Stanford Research 860 lock-in amplifier operating in dual-reference mode to remove the need of a separate demodulation step described previously. 
The resistance is measured simultaneously by a second lock-in amplifier attached to the same voltage contacts detecting the voltage at the same frequency as the excitation current through the sample.
All elastoresistance signals so measured were frequency independent, indicating that the elastoresistance response is intrinsic and not due to elastocaloric heating from the sample or strain device.
The elastocaloric effect in \sample was measured by affixing a Type E chromel-constantan thermocouple onto the region in the middle of the sample, where the strain is expected to be approximately uniform, using a thin layer of AngstromBond AB9110LV epoxy. 
The temperature oscillations at the same frequency as the strain were measured with a lock-in technique.

\begin{figure}
\includegraphics[width=\linewidth]{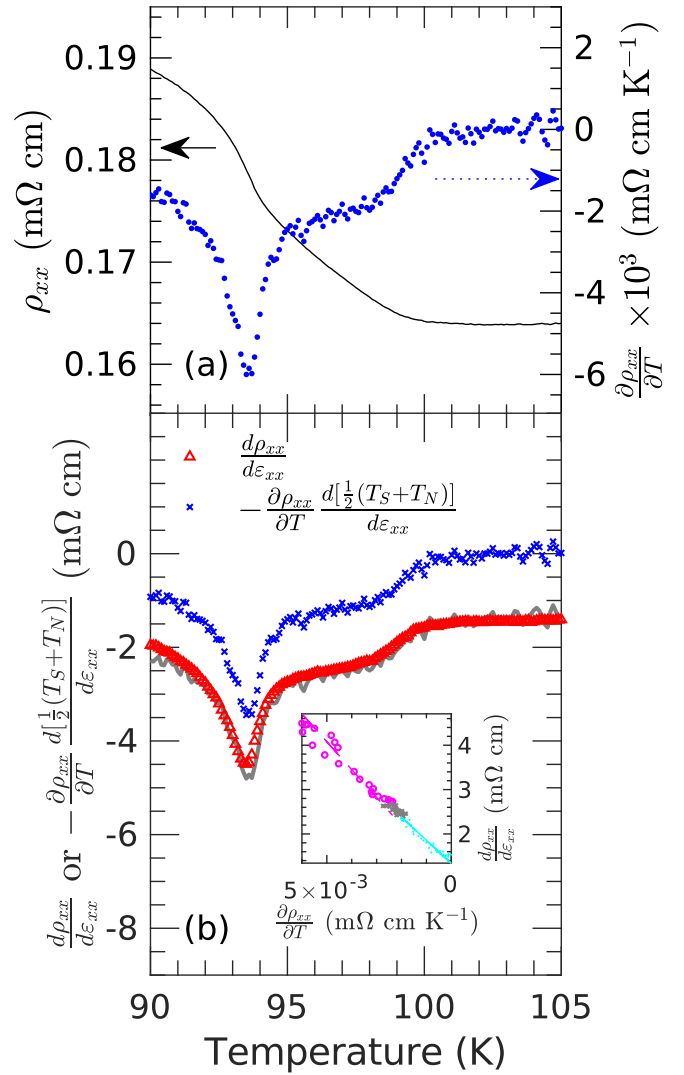}
\caption{(a) Resistivity of sample, plotted as a line against the left axis, and the temperature derivative shown against the right axis. 
(b) Comparison of $-\frac{\partial\rho}{\partial T}\frac{d[\frac{1}{2}(T_S+T_N)]}{d\varepsilon_{xx}}$ and $-\frac{d\rho_{xx}}{d\varepsilon_{xx}}$, motivated by discussion in the text. 
The inset shows the changes in $\frac{\partial\rho_{xx}}{\partial T}$ and $\frac{d\rho_{xx}}{\partial\varepsilon_{xx}}$. Cyan markers represent temperatures in a window of $\pm2$~K around the structural transition; magenta markers represent temperatures in a window of $\pm2$~K around the \neel transition, and grey markers represent temperates outside both windows.
Straight lines show best fits using published values of $\frac{dT_N}{d\varepsilon_{xx}}$ (dashed magenta line) or $\frac{dT_S}{d\varepsilon_{xx}}$(solid cyan line)over the corresponding temperature windows.\cite{IkedaSymmetric2018} 
For the sake of comparison, the trace for $-\frac{\partial\rho}{\partial T}\frac{d[\frac{1}{2}(T_S+T_N)]}{d\varepsilon_{xx}}$ is reproduced with an offset of -1.36 m$\Omega$~cm as a solid gray line (behind the red data points). \label{fig:res}}
\end{figure}

The relation between the elastoresistivity and resistivity is shown in \cref{fig:res}. Inspection reveals a close similarity between the temperature derivative of the resistivity $\partial\rho_{xx}/\partial{T}$ (blue symbols in \cref{fig:res}\,(a)) and the elastoresistivity  $\partial\rho_{xx}/\partial\varepsilon_{xx}$ (red symbols in \cref{fig:res}\,(b)). Motivated by this observation, we scale  $\partial\rho_{xx}/\partial{T}$ by the average value of the strain-induced change in transition temperature $\frac{d[\frac{1}{2}(T_{S}+T_N)]}{d \varepsilon_{xx}} = -575$\,K and plot $\frac{\partial\rho_{xx}}{\partial{T}} \frac{d[\frac{1}{2}[T_{S}+T_N]}{d \varepsilon_{xx}}$(blue crosses in \cref{fig:res} (b)) alongside  $\frac{d\rho_{xx}}{d\varepsilon_{xx}}$. Scaled this way, the data lie almost on top of each other when a constant offset is included (grey line). This paper argues that this correspondence is not accidental.



Though this agreement between the temperature and strain derivatives of the resistivity is striking, we argue it arises naturally from the strain dependence of the critical temperature for each of \neel and nematic order. 
Generally, thermodynamic and transport properties show anomalous behavior on approach to such a strain dependent transition temperature. 
Furthermore, strain perturbations of the type considered here do not change the universality class of the phase transition, so the functional forms of these critical anomalies remain approximately the same in the presence of a strain perturbation.
\footnote{The strains induced in the material belong to either the \aogs or \bogs representations. Both preserve the secondary mirror symmetries broken by the nematic order, and the translational symmetry broken by antiferromagnetic order, and therefore preserve the 3d Ising universality class at each transition.}
This motivates considering anomalous behavior in thermodynamic and transport properties as functions not of temperature and strain separately, but as functions of $T-T_i(\varepsilon)$, where $T_i$ represents $T_s$ or $T_N$ for each of the phase transitions. 

The electronic scattering mechanisms are treated as independent and additive, so the diagonal components of the resistivity can be approximated as having a critical component $\rho^{c}$ and non-critical part $\rho^{(0)}$ satisfying
\begin{equation}
\rho_{xx}(T,\varepsilon_{xx}) \approx \rho_{xx}^{(0)}(T,\varepsilon_{xx}) + \rho^{(c)}_{xx}(T-T_i(\varepsilon_{xx}))
\end{equation}
for $i=S,N$ near each of the nematic and \neel transitions, respectively. This gives rise to an elastoresistive effect 
\begin{equation}
\frac{d\rho^{(c)}_{xx}}{d\varepsilon_{xx}} = -\frac{\partial\rho^{(c)}_{xx}}{\partial T} \frac{dT_i}{d\varepsilon_{xx}}.\label{eranomaly}
\end{equation}
If $\rho^{0}$ has a gradual temperature dependence, the above relation manifests in the derivatives of the total resistivity, as is evident in \cref{fig:res}, and furthermore wherever the Fisher-Langer relation holds, it follows that this elastoresistive response tracks the specific heat.
Motivated by \cref{eranomaly}, we plot in the inset to \cref{fig:res} $\frac{d\rho_{xx}}{d\varepsilon_{xx}}$ versus $\frac{\partial\rho_{xx}}{\partial{T}}$, and overlay linear best fits for the fixed proportionality constants of $\partial T_{i}/\partial \varepsilon_{xx}$ in the proximity of each phase transition. As can be seen the data are consistent with these estimates, up to systematic effects depending on the region around $T_S$ and $T_N$ where the fit is performed.

\begin{figure}
\includegraphics[width=0.9\linewidth]{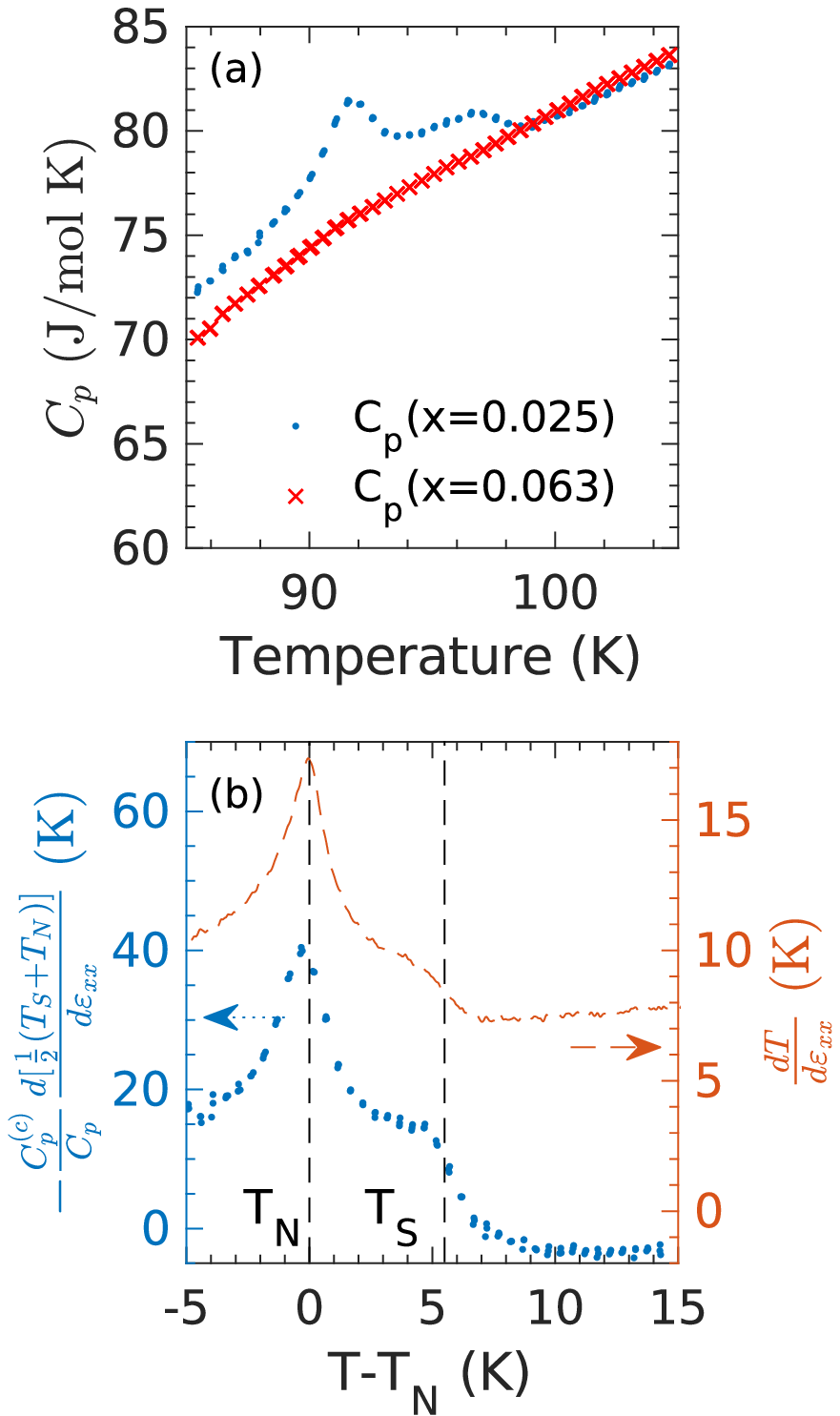}
\caption{(a) The specific heat of \sample \, from ref. \onlinecite{Chu_Determination_2009}. The estimate of the heat capacity anomaly is obtained by subtracting the heat capacity of a sample of \refns, which remains tetragonal and non-magnetic down to T=0. 
(b) Arguments in the text lead to \cref{eceffect}, according to which the critical part of elastocaloric effect (dashed line, right axis) should be exactly equal to the rescaled specific heat anomaly (dots, left axis). Both quantities show similar features at both $T_S$ and $T_N$, however as discussed in the text, an amplitude difference between these two traces arises from the strain apparatus: only part of the sample is strained and the sample heats the surrounding glue. \label{fig:cp}}
\end{figure}

The same model can also be applied to studies of the heat capacity and elastocaloric effect shown in \cref{fig:cp}. 
As before, the energy $U$ of the system is separated into a critical component $U^{(c)}$ and non-critical component $U^{(0)}$. 
As the stress is varied, the critical part of internal energy follows
\begin{equation}
\frac{d U^{(c)}}{d\varepsilon_{xx}} = - C_p^{(c,i)} \frac{dT_i}{d\varepsilon_{xx}},\label{uanomaly}
\end{equation}
where $ C^{(c,i)}_p=\frac{\partial U^{(c)}}{\partial T}$ is the anomaly in the specific heat near the phase transition at $T_i$ and the total specific heat is $C_p$.
As the critical degrees of freedom equilibrate with the other degrees of freedom, the elastocaloric temperature change will be
\begin{equation}
\frac{dT}{d\varepsilon_{xx}} = -\frac{C_p^{(c,i)}}{C_p}\frac{dT_i}{d\varepsilon_{xx}}+\ldots,\label{eceffect}
\end{equation}
neglecting for strain effects on the non-critical degrees of freedom.

\Cref{fig:cp}\,(a) shows the specific heat anomalies, $C_p^{(c,i)}$  of a sample of \sample  obtained by Chu, et al. in ref. \onlinecite{Chu_Determination_2009} by subtracting a ``background" contribution taken to be the measured specific heat of \refns, for which both transitions are absent.
Presumably, these measurements reflect the fact that one transition occurs below and one transition occurs above the respective upper critical dimension, thus the specific heat anomalies consist of a pronounced peak at $T_N$ and a mean-field-like step at $T_S$.  
In both cases, the features are somewhat broadened, possibly reflecting the effect of quenched disorder.
There is a clear correspondence in both the shape of the peak at $T_N$ and the mean-field-like step in both specific heat and elastocaloric coefficient right near $T_S$, though the plateau between transitions is less pronounced in the measurement of the elastocaloric coefficient. 

The difference in magnitude in \cref{fig:cp}\,(b),  approximately a factor of 5, is understood to result from (i) the unstrained parts of the sample\footnote{Based on finite element simulations in Ref. \onlinecite{IkedaSymmetric2018}, it is estimated that effectively only 56\,\% of the sample are strained.} and (ii) the glue attaching the sample to the strain apparatus.
\footnote{The glue between the sample and the bottom mounting plates, which presumably is the part of the mounting glue that is heated the most by the EC effect in the sample, roughly doubles the total sample heat capacity. The thermocouple adds about 2\% to the total sample heat capacity.} 
Most importantly, it should be noted that the observed elastocaloric effect in the sample cannot stem from self heating effects,\footnote{Self heating effects within the PZT actuators would cause a temperature oscillation at double the strain frequency} nor from any elastocaloric effect within the PZT stacks.
\footnote{Elastocaloric effect within the PZT stacks might cause a roughly temperature independent background signal but can not create signatures corresponding to the thermodynamic properties of the sample under investigation. Furthermore, the background signal has been checked to be small by separate temperature measurements on the cell body.} 
Furthermore, our detailed characterization of the experiment\footnote{A manuscript explaining details of the measurement technique will be presented elsewhere.} shows that variation of the thermal properties of the sample and the glue together produce  temperature dependence of the thermal transfer function less than 5\,\% within the temperature window under investigation and for the strain frequency chosen. 
The temperature dependence of the elastocaloric effect signal thus purely reflects the thermodynamic signatures of the sample studied. 


In conclusion, this work demonstrates elastocaloric and elastoresistive anomalies in \sample which track the specific heat and temperature derivative of the resistivity.
Rather than using strain of the same symmetry as the order parameters, as has been used for measurements of thermodynamic suseptibilities,\cite{chu2012divergent,kuo2013measurement,ShapiroSymmetry2015} these effects are realized by inducing strain belonging to different irreducible representations from the order parameter. 
The similarity of these four quantities, shown in \cref{fig:all}, is consistent with a simple picture with just two rules for the critical parts of the resistivity and energy: they satisfy the Fisher-Langer relation and are functions of $T-T_i(\varepsilon)$ close to a phase transition at $T_i$. 
\footnote{For \sample, $i=S$ for the nematic/structural phase transition and, and $i=N$ for the \neel transition.}

Finally, we note that even under circumstances in which the Fisher-Langer relation does not hold, and even for other types of order parameters than those studied here, `isotropic' strain is a ubiquitous tuning parameter to induce the elastocaloric and elastoresistive effects.
Strain which breaks no additional symmetries always has a linear effect on the transition temperature, and therefore results in an elastocaloric effect tracking the specific heat and an elastoresistivity tracking $\frac{d\rho}{dT}$, so strain based techniques like this can be widely applied to extract $\frac{dT_i}{d\varepsilon}$ for any phase transition, or they may provide additional means\cite{Pasler3DXY1998} to extract the critical part of specific heat anomalies.

We acknowledge helpful feedback on our ideas and the manuscript from S. A. Kivelson and T. Worasaran, and further acknowledge J.-H. Chu for providing the specific heat data originally published in ref. \onlinecite{Chu_Determination_2009}.
A.T.H. and J.C.P. are supported by a NSF Graduate Research Fellowship under grant DGE-114747. 
J.C.P. is also supported by a Gabilan Stanford Graduate Fellowship. 
M.S.I. and P.W. were partially supported by the Gordon and Betty Moore Foundations EPiQS Initiative through grant GBMF4414.
This work was supported by the Department of Energy, Office of Basic Energy Sciences, under contract no. DE-AC02-76SF00515.

\bibstyle{aps}
\bibliography{references}

\end{document}